\documentclass[superscriptaddress,twocolumn,showpacs,preprintnumbers,amsmath,amssymb,prb]{revtex4}

\usepackage{graphicx}
\usepackage{dcolumn}
\usepackage{bm}

\begin{document}

\preprint{}

\title{Specific heat of the $S = 1$ spin-dimer antiferromagnet Ba$_3$Mn$_2$O$_8$
in high magnetic fields}

\author{H. Tsujii}
\affiliation{Department of Physics, University of Florida, PO Box
118440, Gainesville, Florida 32611-8440}
\affiliation{RIKEN, Wako, Saitama 351-0198, Japan}
\author{B. Andraka}
\affiliation{Department of Physics, University of Florida, PO Box
118440, Gainesville, Florida 32611-8440}
\author{M. Uchida}
\affiliation{Department of Physics, Tokyo Institute of Technology,
Meguro-ku, Tokyo 152-8551, Japan}
\author{H. Tanaka}
\affiliation{Research Center for Low Temperature Physics,
Tokyo Institute of Technology, Meguro-ku, Tokyo 152-8551, Japan}
\author{Y. Takano}
\affiliation{Department of Physics, University of Florida, PO Box
118440, Gainesville, Florida 32611-8440}

\date{\today}

\begin{abstract}
We have measured the specific heat of the coupled spin-dimer
antiferromagnet Ba$_3$Mn$_2$O$_8$ to 50~mK in temperature and to 
29~T in the magnetic field.
The experiment extends to the midpoint of the field region (25.9~T
$\leq H \leq$ 32.3~T) of the magnetization plateau at 1/2 of the
saturation magnetization, and reveals the presence of three ordered
phases in the field region between that of the magnetization plateau
and the low-field spin-liquid region. The exponent of the phase
boundary with the thermally disordered region is smaller than the
theoretical value based on the Bose-Einstein condensation of spin
triplets.
At zero field and 29~T, the specific-heat data show gapped behaviors
characteristic of spin liquids.  The zero-field data indicate that
the gapped triplet excitations form two levels
whose energies differ by nearly a factor of two.
At least the lower level is well localized.  The data at 29~T
reveal that the low-lying excitations at the magnetization plateau
are weakly delocalized.
\end{abstract}

\pacs{75.30.Kz, 75.40.Cx, 75.50.Ee}
\maketitle

\section{INTRODUCTION}

An interacting array of spin dimers, each consisting of two
antiferromagnetically coupled spins, has a natural tendency to
exhibit plateaus in the magnetization curve $M(H)$. This is obvious
if one considers such a system of $S = 1$ spins. In this case, the
ground state of an isolated dimer is a spin singlet, the lowest
excited state is a spin triplet, and the highest energy state is a
quintet. An increasing magnetic field will cause the ground state to
change from the singlet to the $S_z = +1$ triplet, and then to the
$S_z = +2$ quintet, resulting in two steps in $M(H)$. A
weakly-coupled array of $S = 1$ spin dimers will maintain this
progression of the ground state and will exhibit a magnetization
plateau at $m  = 1/2$, where $m = M/M_{sat}$ is the magnetization
expressed as a fraction of the saturation value.

For one-dimensional antiferromagnets, Oshikawa, Yamanaka, and
Affleck \cite{Oshikawa97} have predicted that $M(H)$ can have a
plateau when the condition $nS(1-m)=$~integer is satisfied. Here $n$
is the period of the ground state. These authors have argued that
the magnetization plateaus are generalized Haldane states with
non-zero magnetization, implying that the transverse spin components
at each plateau form a quantum spin liquid quite like that of the
usual $m=0$ Haldane state. The Oshikawa-Yamanaka-Affleck theorem has
been generalized to higher dimensions by Oshikawa\cite{Oshikawa00}
and refined by other authors.\cite{Misguich}

Plateaus at fractional values of saturation magnetization have been
observed in Cu$^{2+}$-based $S= 1/2$ compounds, such as the
three-dimensional (3D) spin-dimer antiferromagnet NH$_4$CuCl$_3$
\cite{Shiramura}, the quasi-2D orthogonal-dimer antiferromagnet
Sr$_2$Cu(BO$_3$)$_2$ \cite{Kageyama}, and the frustrated quasi-2D
spin system Cs$_2$CuBr$_4$.\cite{Ono04} NH$_4$CuCl$_3$ shows distinct
plateaus at $m=1/4$ and 3/4, Sr$_2$Cu(BO$_3$)$_2$ at $m=1/8$, 1/4,
and 1/3, and Cs$_2$CuBr$_4$ at $m=1/3$ and 2/3.  In addition, plateaus at $m=
1/2$ have been observed in Ni$^{2+}$-based $S=1$ Heisenberg
antiferromagnets with bond alternations. \cite{Narumi98,Narumi04}

Ba$_3$Mn$_2$O$_8$ is a new $S=1$ spin-dimer antiferromagnet, in
which the inter-dimer coupling is expected on a structural ground to
be three dimensional. The magnetization of this compound has been
measured at 0.65 K as a function of the magnetic field up to 50 T in
a pulsed magnet.\cite{Uchida02} The data exhibit a low-field $m = 0$
region and an $m =$ 1/2 plateau, which occurs at around 29 T. In this
paper, we report a determination of the magnetic phase diagram
over the entire field range below the $m =$ 1/2 plateau, by means of
specific heat measurements performed down to 50 mK in temperature.
We also examine the nature of the excitations at zero field
and at the $m=$ 1/2 plateau.

Ba$_3$Mn$_2$O$_8$ has a trigonal structure of the space group
$R{\bar 3}m$ with the lattice parameters $a = 5.711$ \AA \ and $c =
21.444$ \AA. The manganese ions are in the unusual Mn$^{5+}$ state,
which makes $S = 1$. To date, all experiments including the present
work have been performed on powder samples, since no single crystal
has been successfully grown to our knowledge. The magnetization
measured at 0.65~K is essentially zero below $H_{c1}=$ 9.2~T, where
it starts to increase linearly with the magnetic field. \cite{Uchida02}
The $m=$ 1/2 plateau extends from $H_{c2}=$ 25.9~T to $H_{c3}=$
32.3~T, and the saturation magnetization is reached at
$H_{sat}\approx$ 48~T. The zero-field gap $\Delta$ has been
estimated from $H_{c1}$ to be $\Delta = g \mu_{\rm B}H_{c1} =$ 12.2
K. From the overall shape of the magnetization curve, the strengths
of the intra-dimer and inter-dimer exchanges have been determined to
be $J_0 =$ 17.4~K and $J_1+2(J_2+J_3) =$ 8.3~K, respectively, where
$J_0$, $J_1$, $J_2$, and $J_3$ are the first-, second-, third-, and
fourth-nearest-neighbor interactions, respectively. The edges of the
magnetization plateau are sharp in spite of the powder form of the
sample, indicating an isotropy of the spin hamiltonian. In electron
spin resonance, the absorption spectrum has a single Lorentzian peak
with a narrow linewidth, indicating the smallness of any anisotropic
exchange interaction and the good isotropy of the $g$ factor, which
is determined to be $g=$ 1.98. \cite{Uchida02,Uchida01}  This $g$
value has been used for the estimate of the gap $\Delta$. The
single-ion anisotropy term is estimated to be only $D\sim 0.02\bar
J$ from the linewidth, where $\bar J$ is the average of the exchange
interactions. \cite{Uchida02}

\section{EXPERIMENTAL METHODS}

The powder sample of Ba$_3$Mn$_2$O$_8$ used in the present work was
synthesized by the method described in Ref.~\onlinecite{Weller},
with an extended sintering time of 150~h to improve the sample
quality. \cite{Uchida02}  Specific heat measurements to 18~T were
performed in a superconducting magnet at temperatures down to 50~mK
using a dilution refrigerator. The relaxation calorimeter for this
setup has been described in Ref.~\onlinecite{Tsujii03}. For
measurements from 19.5~T to 29~T, another relaxation calorimeter
with a built-in $^3$He refrigerator was used in a resistive Bitter magnet.

Powder samples present a challenge to relaxation calorimetry at
temperatures typically below 1~K because of the difficulty in
obtaining a good thermal contact between the powder grains and the
calorimeter.  To solve this problem, we mixed the sample with a
small amount of silver paint \cite{Arzerite} and sandwiched it
between two 0.13~mm-thick pieces of sapphire to form a thin, uniform
layer. The bottom piece of sapphire was attached with the Wakefield
compound \cite{Wakefield} to the calorimeter platform. Three samples
weighing 0.15~mg, 1.08~mg, and 7.62~mg were used to cover different
temperature and field regions.

\begin{figure}[tbp]
\begin{center}\leavevmode
\includegraphics[width=0.8\linewidth]{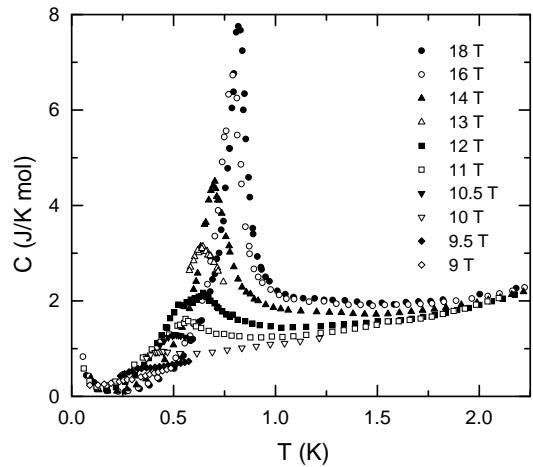}
\caption{ Temperature dependence of the specific heat of
Ba$_3$Mn$_2$O$_8$ in constant magnetic fields between 9~T and 18~T.
The nuclear contribution of $^{55}$Mn is visible at temperatures
below 0.2~K. }\label{fig1}\end{center}\end{figure}

\section{EXPERIMENTAL RESULTS AND DISCUSSIONS}

\subsection{Magnetic phase diagram}

The temperature dependence of the specific heat is shown in Fig.~1
for fields between 9~T and 18~T.  In fields higher than 13~T, a well
defined peak appears in the specific heat below 1~K, signalling a
phase transition. The sharpness of the peak indicates a high quality
of the sample despite its powder form. An additional peak which
indicates a second transition appears at 12~T. In fields lower than
this value, the specific heat anomaly becomes weak, and we can
identify only one peak with confidence.  No peak is found at 9~T.

The specific heat in magnetic fields higher than 18~T, measured in
the resistive Bitter magnet, is shown in Fig.~2. The peak in the
specific heat is sharp up to 22~T. The 24~T data are very similar
to the 12~T data, except that there are two shoulders instead of two
peaks.  At 25~T, the anomaly becomes small, and only one broad peak
can be identified.

\begin{figure}[btp]
\begin{center}\leavevmode
\includegraphics[width=0.8\linewidth]{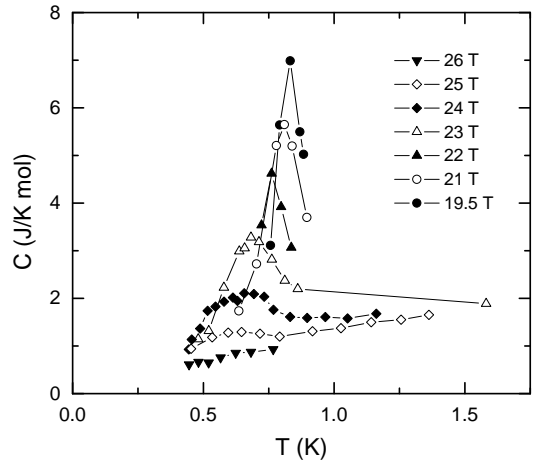}
\caption{ Specific heat of Ba$_3$Mn$_2$O$_8$ in constant magnetic
fields ranging from 19.5 T to 26 T. The lines are guides to the eye.
}\label{fig2}\end{center}\end{figure}

To further explore the field region below 13~T, we have measured the
specific heat as a function of the magnetic field. In these
measurements, a constant electric current was applied to the heater
of the thermal reservoir, allowing the temperature to rise
monotonically with increasing field as dictated by the
magnetoresistance of the heater. As seen in Fig.~3, the phase
transitions are observed as a peak and shoulder.

\begin{figure}[btp]
\begin{center}\leavevmode
\includegraphics[width=0.65\linewidth]{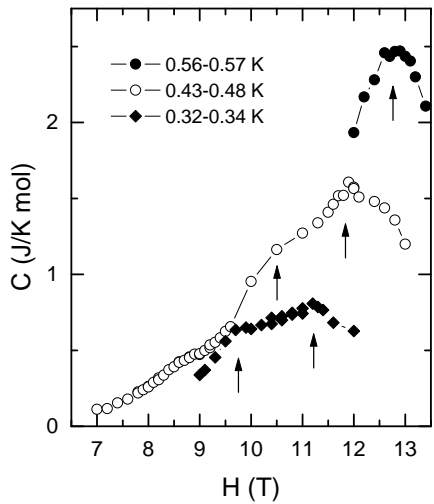}
\caption{ Magnetic-field dependence of the specific heat of
Ba$_3$Mn$_2$O$_8$. The transition points are indicated by arrows.
The lines are guides to the eye.
}\label{fig3}\end{center}\end{figure}

The magnetic phase diagram determined from the positions of the
specific-heat peaks and shoulders in Figs.~1 through 3 is given in
Fig.~4. As seen in the figure, the transition temperatures obtained
in the magnetic-field scans shown in Fig.~3 agree with those
obtained in the temperature scans shown in Fig.~1. The phase diagram
contains three ordered phases in the investigated field region.
In addition to the $H_{c1}$ and $H_{c2}$ observed in the
magnetization data\cite{Uchida02}, there are two critical fields at
the intermediate fields of about 11~T and 24.5~T.
Figure~4 is somewhat reminiscent of the magnetic phase diagram of
CsFeBr$_3$, \cite{Tanaka01} which is an $S=1$ antiferromagnet on a
triangular lattice with a singlet ground state. In this material,
$H_c$ of the field-induced ordering at zero temperature splits into
two branches as the temperature is raised, with the separation
between the two branches widening with increasing temperature.
\cite{Nakamura03} In contrast, Fig.~4 suggests two branches of
either $H_{c1}$ or $H_{c2}$ that are separate at zero temperature
merging into one as the temperature is raised.

\begin{figure}[btp]
\begin{center}\leavevmode
\includegraphics[width=0.8\linewidth]{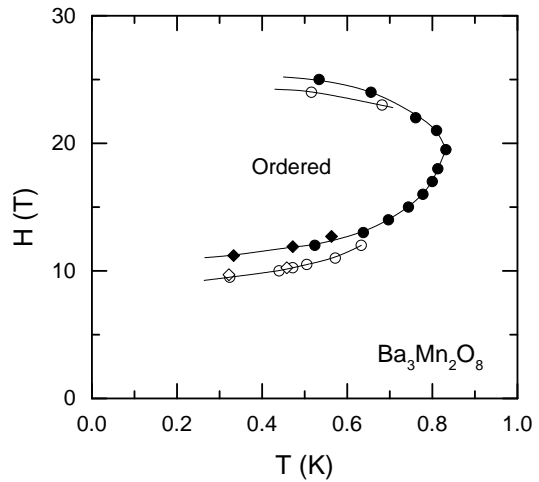}
\caption{ Magnetic phase diagram of Ba$_3$Mn$_2$O$_8$. The circles
are the positions of the peaks and shoulders in the specific heat at 
constant magnetic fields.  The diamonds are the positions of the peaks 
and shoulders in the specific heat measured as a function of the magnetic 
field, while the current for the reservoir heater was held constant. 
The lines are guides to the eye. }\label{fig4}\end{center}\end{figure}

The Bose-Einstein condensation of triplets is believed to be a valid
description of field-induced long-range ordering in all gapped
antiferromagnets. \cite{Tsvelik,Nikuni} According to the
Bose-Einstein condensation theory, the phase boundary in the
temperature \emph{vs} field plot obeys a power law $T_{\rm c}
\propto (H - H_{\rm c} )^\alpha$. The field-induced ordering in the
$S=1/2$ spin-dimer material TlCuCl$_3$ has been interpreted in terms
of the Bose-Einstein condensation. \cite{Nikuni} However, the
exponent $\alpha$ of 0.50 determined for this material is smaller
than the theoretical value of 2/3 obtained with a Hartree-Fock
approximation.

We have obtained $H_{c1}=9.04 \pm 0.15$~T and $\alpha=0.39 \pm 0.06$
by fitting the data of the phase boundary between the disordered
phase and the ordered phase below 0.58~K. The zero-temperature
critical field $H_{c1}$ is in good agreement with 9.2~T obtained by
the magnetization curve at 0.65~K.  Note that the critical fields
determined by a low-temperature magnetization curve $M(H)$ are
generally closer to the zero-field values than the actual transition
fields at the temperature of the measurement. The value for the
exponent $\alpha$ is smaller than 2/3 predicted by the theory.
However, it agrees within the experimental uncertainty with the
quantum Monte-Carlo result $\alpha=0.37 \pm 0.03$ for $S =$ 1/2 spin
dimers with a weak three-dimensional inter-dimer coupling.
\cite{Wessel} It is also close to the values for the $S=1$
spin-chain materials NTENP and NDMAP for the field applied parallel
to the chain direction, $\alpha=0.334\pm0.004$ and $0.34\pm0.02$ for
NTENP\cite{Tateiwa} and NDMAP, \cite{Tsujii05} respectively.
Recently, careful analyses near the zero-temperature critical field
$H_{c1}$ have shown that the exponent $\alpha$ approaches 2/3, as
the temperature region of the fits is narrowed, for both the quantum
Monte-Carlo data \cite{Nohadani} and the experimental data \cite{Shindo} 
for TlCuCl$_3$. A similar analysis has yielded $\alpha = 0.63\pm0.03$ for
BaCuSi$_2$O$_6$, an $S=1/2$ spin-dimer compound with predominantly
two-dimensional inter-dimer exchanges. \cite{Sebastian}
It remains to be seen, however, whether the
Bose-Einstein condensation theory can explain the field-induced
magnetic ordering in all gapped antiferromagnets.

\subsection{Gapped excitations at zero field and at the magnetization plateau}

In the low-field $m = 0$ region, as well as at the $m = 1/2$ magnetization
plateau, the specific heat is expected to exhibit a gapped behavior.  The
data at zero field and 29~T, the midpoint of the plateau, are shown in
Fig.~5. At zero field, the low-temperature
behavior of the data indicates the presence of a gap.  
However, the low-temperature specific heats of delocalized gapped
excitations have wrong temperature dependences --- see Eq.~4 for an
example --- which do not agree with the data.
The best fit
of the data below 3.5~K is obtained by using the low-temperature
formula for localized gapped excitations,
\begin{equation}
C=\tilde{n}R(\Delta/T)^{2}e^{-\Delta/T}, 
\end{equation}
where $\tilde{n}$ is the
number of excited states per spin dimer, and $R$ is the gas constant.  
As shown in the inset of Fig.~5, this formula gives the
best fit with an energy gap of $\Delta=12.5\pm0.2$~K, which is in
excellent agreement with 12.2~K estimated from $H_{c1}$. 
In Ba$_3$Mn$_2$O$_8$, the average inter-dimer exchange $J_1+2(J_2+J_3)
=$ 8.3~K is substantial in comparison with the intra-dimer exchange
$J_0 =$ 17.4~K, as we have described.  
It is surprising that, despite such strong inter-dimer exchanges, the triplet 
excitations remain localized instead of becoming highly dispersive.  
The crystal structure \cite{Uchida02} of Ba$_3$Mn$_2$O$_8$ suggests that 
the localization arises from a geometric frustration.

\begin{figure}[btp]
\begin{center}\leavevmode
\includegraphics[width=0.81\linewidth]{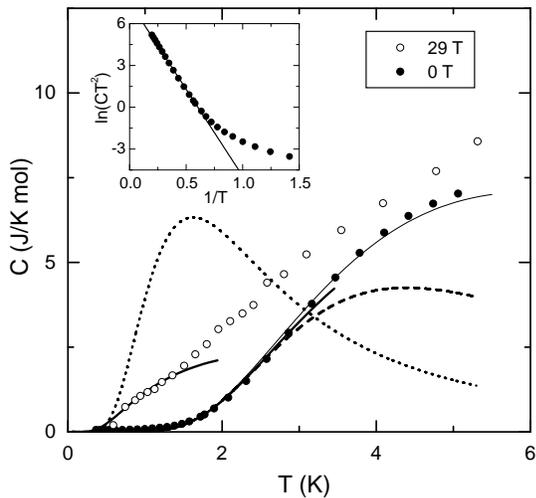}
\caption{ Temperature dependence of the specific heat of
Ba$_3$Mn$_2$O$_8$ at 0~T and 29~T. The thick solid lines are the
low-temperature fits to the data.  The dashed line indicates the
calculated zero-field specific heat due to localized, gapped excitations,
whose parameters are taken from the low-temperature fit. 
The thin solid line is the specific heat due to two kinds of localized 
excitations with two gap energies, as described in the text.
The dotted line for the 29~T data indicates the
specific heat of mutually isolated dimers with an excitation gap of
$\Delta =$ 4.3~K. 
In the inset, $\ln(CT^2)$ of the zero-field data have been plotted against $1/T$
to reveal the localized-excitation behavior, which is given by the
straight line.  The deviation of the data from the line at temperatures
below 1.4~K is due to the phonon contribution and an uncertainty in the
subtraction of the background heat capacity of the silver paint.}
\label{fig5}\end{center}\end{figure}

Another surprising result of the low-temperature fit is that
$\tilde{n}$ is only $1.44 \pm 0.04$, which is very close to 1/2 of $\tilde{n}= 3$
expected for the triplets.  The dashed line in the main panel of the figure shows the
contribution of local excitations for
the entire temperature range of the data, according to the complete
expression 
\begin{equation}
C=\tilde{n}R\frac{(\Delta/T)^2 e^{-\Delta/T}}{(1+\tilde{n} e^{-\Delta/T})^2},
\end{equation}
for $\tilde{n}=1.44$ and $\Delta = 12.5$~K found 
by the low-temperature fit.  
Above 3.5~K, the specific heat is substantially larger than can be
accounted for  by these excitations, 
strongly suggesting that the excess arises from the ``missing" triplets 
whose gap is considerably larger than 12.5~K.  Indeed, 
as shown by the thin solid line, the data over the entire
temperature range can be well described
by 1/2 of the triplets at $\Delta=12.5\pm0.2$~K and the other 1/2 
at $\Delta'=24.5\pm0.5$~K: 
\begin{widetext}
\begin{equation}
C=\tilde{n}R\frac{(\Delta/T)^2 e^{-\Delta/T}+(\Delta'/T)^2 e^{-\Delta'/T} + 
\tilde{n} (\Delta'/T-\Delta/T)^2 e^{-(\Delta+\Delta')/T}}{[1+\tilde{n} (e^{-\Delta/T}+e^{-\Delta'/T})]^2},
\end{equation}
\end{widetext}
where $\tilde{n}=1.5$.  We find $\Delta'=2 \Delta$ within the combined
uncertainties of the fitting parameters.  Although Eq.~3 assumes that the
excitations at $\Delta'$ are localized like those at $\Delta$, the data do not
tell whether this is the case, since the excitations at $\Delta$ dominates
the low-temperature behavior which distinguishes localized excitations
from dispersive excitations.

The field of 29~T is located at the midpoint of the $m=1/2$ magnetization
plateau within the uncertainty of the magnetization data. As
expected, no sign of ordering is observed in the specific-heat data
at this field, indicating that the plateau region is a spin-liquid
phase, although there is a substantial scatter at high temperatures.
At the midpoint of an $m=1/2$ magnetization plateau, mutually
isolated $S=1$ dimers will have the lowest excited level containing
two states, a singlet and an $S_z=2$ quintet. The gap for the level
of the collective excitations analogous to such single-dimer states
can be estimated from the width of the magnetization plateau to be
$\Delta = g\mu_{\rm B}(H_{c3}-H_{c2})/2=$ 4.3~K. 
The specific heat due to localized gapped excitations with $\tilde{n}=$
2 and $\Delta=$ 4.3~K, shown by the dotted line, completely fails to
describe the data at 29~T. The comparison of the line with the data
in fact indicates that the energy of the excitations has a wide
distribution, suggesting that their dispersion cannot be neglected
in comparison with the gap size.

By following the method described by Troyer \emph{et al.}
\cite{Troyer} for one-dimensional excitations, one can derive the low-temperature expression
\begin{equation}
C=n_{m}R\frac{\Delta^{2}}{(4\pi a)^{3/2}T^{1/2}}e^{-\Delta/T},
\end{equation}
for the specific heat due to \textit{three-dimensional} excitations whose gapped
dispersion is $\epsilon=\Delta+a(k-k_0)^2$. Here $n_{m}$ is the
number of the gapped modes. This expression gives the best fit to
the data below 1.5~K as shown in Fig.~5, with $\Delta=1.9\pm0.1$~K,
which is considerably smaller than 4.3~K anticipated by the simple
argument given above. The strength of the dispersion can be
estimated from the fitting parameter $a/n_{m}^{2/3}=0.20\pm0.01$~K.
Regardless of the number of modes $n_{m}$, which we do not know at
the moment, this is a rather small value, indicating that the
low-lying excitations at the $m=1/2$ magnetization plateau are
nearly localized, similar to the low-lying excitations at zero
field.  The measured specific heat continues to rise above 1.5~K,
where the low-temperature approximation is invalid, revealing the 
presence of at least another low-lying mode whose energy is on the
order of 10~K. This observation, together with the experimentally
obtained value for $\Delta$ being considerably smaller than the
value estimated from the plateau width, suggests that the levels of
the singlet and the $S_z=2$ quintet do not cross each other at the
midpoint of the magnetization plateau.

\section{CONCLUSIONS}

In this work, we have determined the magnetic phase diagram of Ba$_3$Mn$_2$O$_8$ 
by using specific-heat measurements to 50~mK in temperature and
to 29~T in the magnetic field.
The phase diagram reveals the existence of three ordered phases
between the low-field spin-liquid region and the $m=1/2$
magnetization plateau.
We have found that the exponent for the transition temperature of
the field-induced ordering in the investigated temperature region is
clearly less than 2/3 expected by the Bose-Einstein condensation theory.  

In the spin-liquid phase at zero field, 
the specific heat over the entire temperature range is well
accounted for by two excitation levels, each probably containing the same number of states.  
The gap $\Delta$ for the lower level agrees well with the value estimated from
$H_{c1}$, and the excitations at this level are well localized.
The existence of the higher level at about $2\Delta$ requires direct 
confirmation by ESR or inelastic neutron scattering.

In the spin-liquid phase at the $m=1/2$ magnetization
plateau, the low-lying excitations are delocalized, but the dispersion is 
weak. Contrary to
a naive expectation, the two collective modes arising from the
singlet and the $S_z=2$ quintet of a single dimer do not cross each
other at the midpoint of the plateau.
We hope this finding will stimulate further theoretical work on 
three-dimensionally coupled spin dimers.  In addition, it will be useful to
perform an ESR experiment to map the excitation energies in the magnetization 
plateau region as a function of field.

\begin{acknowledgments}
We thank M.~Takigawa for useful information on the hyperfine field
of $^{55}$Mn, and F.~C.~McDonald,~Jr., T.~P.~Murphy, E.~C.~Palm, and
C.~R.~Rotundu for assistance. This work was supported by the NSF
through DMR-9802050 and the DOE under Grant No. DE-FG02-99ER45748. A
portion of it was performed at the National High Magnetic Field
Laboratory, which is supported by NSF Cooperative Agreement No.
DMR-0084173 and by the State of Florida.
\end{acknowledgments}

\end{document}